\documentclass[%
 reprint,
nofootinbib,
 amsmath,amssymb,
 aps,
pra,
]{revtex4-2}
\usepackage{hyperref}
\usepackage{siunitx}
\usepackage{algorithm}
\usepackage{comment}

\usepackage{algorithmicx}
\usepackage{algpseudocode}
\usepackage{xcolor}
\hypersetup{colorlinks,linkcolor={red!50!black},citecolor={blue!50!black},urlcolor={blue!80!black}}
\usepackage{mathtools, stmaryrd}
\usepackage{amssymb}
\usepackage{float}
\usepackage{mathrsfs,amsmath}
\usepackage{braket}
\usepackage{mhchem}
\usepackage{graphicx}
\usepackage{dcolumn}
\usepackage{footnote}
\usepackage{bm}
\usepackage{mathtools, stmaryrd}
\usepackage{xparse} \DeclarePairedDelimiterX{\Iintv}[1]{\llbracket}{\rrbracket}{\iintvargs{#1}}
\NewDocumentCommand{\iintvargs}{>{\SplitArgument{1}{,}}m}
{\iintvargsaux#1} %
\NewDocumentCommand{\iintvargsaux}{mm} {#1\mkern1.5mu..\mkern1.5mu#2}

\begin{document}

\preprint{APS/123-QED}

\title{Ultra-precise holographic optical tweezers array}

\author{Y. T. Chew\textsuperscript{1,*},
M. Poitrinal\textsuperscript{1,*},
T. Tomita\textsuperscript{1,2},
S. Kitade\textsuperscript{1,3},
J. Mauricio\textsuperscript{1,2},
K. Ohmori\textsuperscript{1,2},
S. de Léséleuc\textsuperscript{1,4}
}

\affiliation{\textsuperscript{1}Institute for Molecular Science, National Institutes of Natural Sciences, Okazaki, Japan}
\affiliation{\textsuperscript{2}The Graduate University for Advanced Studies (SOKENDAI), Okazaki, Japan}
\affiliation{\textsuperscript{3}Nara Institute of Science and Technology (NAIST), Ikoma, Japan}
\affiliation{\textsuperscript{4}RIKEN Center for Quantum Computing (RQC), Wako, Japan}

\date{\today}

\begin{abstract}

Neutral atoms trapped in microscopic optical tweezers have emerged as a growing platform for quantum science. Achieving homogeneity over the tweezers array is an important technical requirement, and our research focuses on improving it for holographic arrays generated with a Spatial Light Modulator (SLM). We present a series of optimization methods to calculate better holograms, fueled by precise measurement schemes. These innovations enable to achieve intensity homogeneity with a relative standard deviation of 0.3\%, shape variations below 0.5\%, and positioning errors within 70~nm. Such ultra-precise holographic optical tweezers arrays allow for the most demanding applications in quantum science with atomic arrays. 
\end{abstract}

\maketitle
\renewcommand{\thefootnote}{\fnsymbol{footnote}}
\setcounter{footnote}{0}
\footnotetext{These authors contributed equally.}

\section{\label{sec:level1} Introduction}
Over the past five decades, optical tweezers have evolved into powerful tools for manipulating microscopic objects~\cite{Ashkin_2000}. These devices, exerting optical force through a finely focused laser beam, enable precise control over the motion of diverse elements, from complexly structured cells~\cite{Ashkin1987_cell} to nanoparticles~\cite{Ashkin_1986} and even individual atoms~\cite{Schlosser2001tweezer}. Advances in optical engineering have significantly broadened their applicability, establishing them as critical tools in probing biochemical processes~\cite{Bustamante2021Mar_bio_review}, managing nanoparticle dynamics within quantum regimes~\cite{Stickler2021Aug_rot_nano_review}, and assembling arrays of ultracold trapped atoms~\cite{Kaufman2021Dec_review}.

Among the diverse applications of optical tweezers, the capability to trap single atoms in optical tweezers arrays forms the primary focus of this work. These atomic arrays have become essential tools for a diverse range of studies, encompassing quantum simulation~\cite{Browaeys2020Feb_review}, computation~\cite{Graham2022,Bluvstein2024Feb}, and metrology~\cite{Madjarov_endres_clock_2019,Eckner2023}. Optical tweezers arrays have been created by adjustable devices such as Spatial Light Modulators (SLMs)~\cite{Bergamini04,Nogrette2014,Kim2016Oct,Barredo2018}, acousto-optic deflectors \cite{kaufman_2014_science, Endres2016Nov} or digital micromirror devices~\cite{Zupancic2016,Wang_whitlock_2020}, or by fixed tools such as intensity masks~\cite{Saffman2022_mask} or microlens arrays~\cite{microlens_2023_prl}. Among these devices, SLMs excel in creating arbitrary geometries with high efficiency, with recent works reporting on scaling up to thousands of tweezers~\cite{manetsch2024tweezer,dmd_jeff_2024,pichard2024rearrangement_pasqal}.

For precise manipulation of the atoms, maintaining homogeneity across the tweezers array is essential. This problem has already been tackled by the community~\cite{LeonardoTweezers,MatsumototoTweezers,MatsumotoTweezers_2,Nogrette2014,TamuraTweezers,KimLukinOptica2019,tweezerbrowaeys2021,ren2024creation_neural_2024}.  Here we report on an upgrade of the hologram optimization method utilizing \textit{in situ} measurement to create arrays which are highly homogeneous in tweezers \textit{intensity}, \textit{shape} and \textit{position}.
This article is organized as follows. Beginning with a brief overview of computer-generated holograms (CGHs) and a simplified formalism of CGHs for arrays of optical tweezers in Section~\ref{sec:level2}, the details of the optimization procedure are explored in the subsequent sections. Section~\ref{sec:intensity} focuses on the intensity homogenization, Section~\ref{sec:level4} on the shape, and Section~\ref{sec:position} on the position. In each section, feedback methods are adopted for the precise optimization of the CGHs, and the relevant \textit{in situ} measurement utilizing the signal from the atoms is introduced and discussed. As a highlight of this work, intensity inhomogeneity of 0.3\%, shape inhomogeneity below 0.5\%, and position inhomogeneity of 70~nm are achieved. Eventually, a conclusion and outlook are provided in Section~\ref{sec:summary}.

\section{\label{sec:level2}DESIGNING HOLOGRAMS}
In this section, we explore the process of generating CGHs and their optimization. This includes a review of the conventional methods for hologram calculation and a simplified interpretation employing elementary phase holograms~\cite{LeonardoTweezers}. After outlining this calculation method, we discuss the sources of inhomogeneities and describe how to correct them using the same procedure for different types of inhomogeneities.

\begin{figure}[]
    \centering
    \includegraphics[width=\columnwidth]{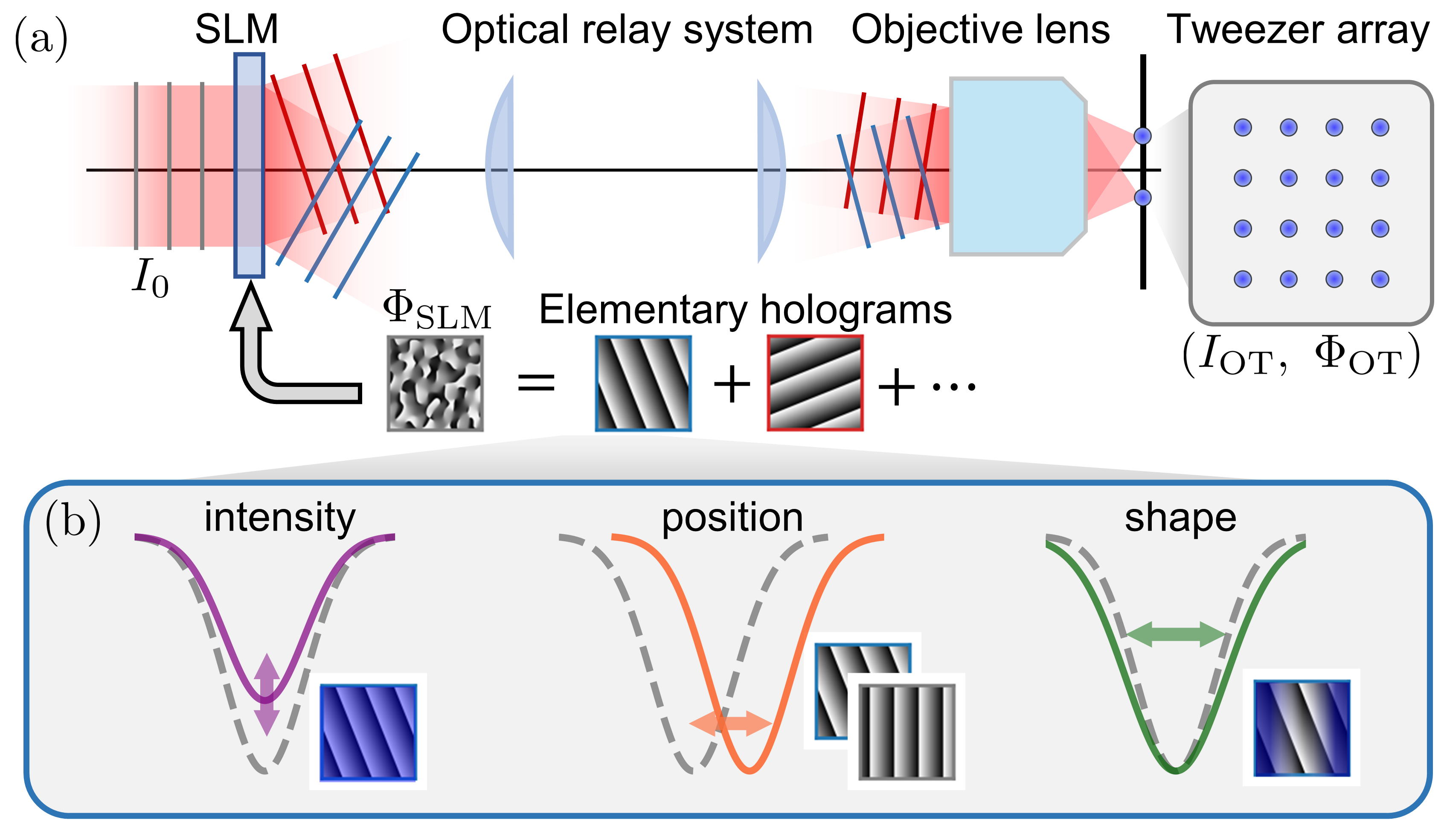}
    \caption{\textbf{Generation of optical tweezers arrays using an SLM}. (a) An SLM is employed to modulate spatially the incoming laser beam and create an array of optical tweezers. The phase of the SLM can be reduced to a superposition of elementary holograms for our specific array. (b) Imperfections lead to intensity, position and shape inhomogeneity between tweezers. Specific modifications of the elementary holograms in the SLM phase calculation can correct these inhomogeneities.  }
    \label{fig1}
\end{figure}

A simple schematic of the system we consider is shown in Fig.~\ref{fig1}(a). A single optical tweezers can be created by focusing a laser beam, of wavelength $\lambda$, with an objective lens of focal $F$. To create an array of $N$ tweezers, we would like to input $N$ \textit{beamlets} with different angles of incidence at the objective entrance. This is achieved in a scalable way by using a hologram imprinted on the SLM. The laser beam is thus first shone on the SLM, which modulates spatially the laser phase with a two-dimensional phase mask $\Phi_{\text{SLM}}$, without modifying its intensity. The diffraction of light from the phase mask creates the array of beamlets (red and blue wavefronts). These are then optically relayed by one (or a series of) telescope imaging the SLM plane into the objective lens, whose magnification $M$ is chosen to match the beam diameter illuminating the SLM with the back aperture of the objective. The objective then forms the array of tweezers in its focal plane, which is simply the diffraction pattern $I_{\rm OT}$ of the phase mask
\begin{equation}
I_{\rm OT} = |\mathscr{F}(A_{\rm SLM})|^2 
\label{eq:ft}
\end{equation}
where $\mathscr{F}$ is the two-dimensional Fourier transform of the complex amplitude of the laser light after the SLM:
\begin{equation}
A_{\rm SLM} = \sqrt{I_{\rm 0}} \exp{(i \Phi_{\rm SLM})}
\end{equation}
and $I_{\rm 0}$ the intensity profile of the incoming laser beam.


Our task is now to find a phase mask $\Phi_{\rm SLM}$ generating a target tweezers intensity pattern $I_{\rm OT}$. This is a well-known problem, called the phase-retrieval problem, which we present again for the sake of pedagogy. The task could be trivially solved by displaying the inverse Fourier transform of $I_{\rm OT}$ on the SLM plane. However, this would require both phase and intensity modulation on the SLM plane, which is not possible with a phase-only spatial modulator. So, in addition to targeting the tweezers pattern $I_{\rm OT}$, we also add the constraint of a fixed $I_{\rm 0}$. The problem is solvable because we are not constraining the phase in the focal plane $\Phi_{\rm OT}$, which is unimportant in our context.


Solving this phase-retrieval problem can be achieved with different approaches. One possibility is to perform a \textit{brute-force} optimization of the hologram~\cite{Harte_slm_opti_2014}, where the value of the phase of each pixel of the SLM is optimized with respect to a cost function. Another approach is based on Iterative Fourier-Transform Algorithm (ITFA), whose most famous example is the Gerchberg-Saxton (GS) algorithm~\cite{GerchbergSaxton1972, Wyrowski_ft_1988}. We will follow this approach, but simplify its formulation with respect to previous works~\cite{Nogrette2014,KimLukinOptica2019}, as we are only considering a rather simple set of target patterns: an arbitrary array of well-separated spots. 



We will now see how to build up a hologram from \textit{elementary} patterns, and then how to optimize it. First, let's consider how to produce a single tweezers displaced from the optical axis by $(x_m,y_m)$, where $m$ will later become the tweezers index. This requires to tilt the incoming laser beam on the SLM, which is achieved with a phase gradient in the direction of the tilt. This elementary phase pattern, represented in the blue and red box of Fig.~\ref{fig1}(a), is described as $\Phi(\bold{r}) = \bold{k_m}\cdot\bold{r} + \theta_m$, where $\bold{r}$ is the coordinates in the SLM plane, $\mathbf{k_m} =  \frac{2\pi M}{\lambda F}(x_m,y_m)$ is the grating wave-vector, and $\theta_m$ is an arbitrary phase, which we will soon discuss. The phase is wrapped in a $2\pi$-range to be displayed on the SLM. 

Secondly, let's \textit{add-up} elementary patterns to make the complete hologram following: 
\begin{align}
A_{\rm SLM} (\mathbf{r)} &= \sum_{m}{f_m(\mathbf{r})e^{i(\mathbf{k_m}\mathbf{r}+\theta_m )}} \label{eq:interference_tweezers}
\\
\Phi_{\rm SLM}(\mathbf{r}) &= {\rm arg}(A_{\rm SLM}(\mathbf{r}))
\end{align}
where the term $f_{m}(\mathbf{r})$ is first set to 1 and will be used later to correct inhomogeneities.
A simple physical picture is obtained by looking at Fig.~\ref{fig1}(a) from right to left, and considering the interference of beamlets `originating' from the array of tweezers, and forming the interference pattern $A_{\rm SLM}$, whose phase argument then gives the phase mask $\Phi_{\rm SLM}$. The apparently complicated hologram forming the tweezers array is simply the interference of many elementary holograms, with different wave-vectors and phases, related to the tweezers position and phase. Mathematically, Eq.(\ref{eq:interference_tweezers}) is the inverse Fourier transform of the tweezers pattern. Thanks to this simple analytical expression in the case of an array of spots, we prefer to write it explicitly.  

Thirdly, and as mentioned earlier, because the SLM only displays the phase information and not the intensity information contained in Eq.(\ref{eq:interference_tweezers}), we need to optimize the choice of $N$ phases $\theta_m$ to produce an interference pattern whose intensity $|A_{\rm SLM}|^2$ is as close as possible to the constraint $I_{\rm 0}$. Deviation from this will decrease the diffraction efficiency of the hologram, as it will not correctly match the incoming light mode to the one corresponding to the tweezers array. A very bad choice of phases $\theta_m$ would be to have them all equal, as the interference of a phased array of spots would create highly-contrasted minima and maxima, very far from the constraint of a quasi-uniform illumination of the SLM. A much better choice is thus to take random phases, which is in fact the appropriate starting point of the Gerchberg-Saxton optimization loop. The GS routine then proceeds as follows: a phase mask is calculated based on Eq.(\ref{eq:interference_tweezers}), then its diffraction pattern is numerically calculated with Eq.(\ref{eq:ft}) with the phase of the tweezers used to proceed to the next iteration. All calculations can be run efficiently on a GPU. This algorithm converges in roughly 10 steps, producing an optimized hologram whose calculated diffraction efficiency is typically 90~\%.  

We have now produced an efficient hologram generating an array of tweezers. However, it is plagued by inhomogeneities. The most discussed in the literature is a tweezers-to-tweezers deviation of the peak intensity, which can be as high as 20~\%. In this work, and as depicted in Fig.~\ref{fig1}(b), we will also consider deviations of the tweezers position and its shape (more precisely, its curvature at the center). These three types of inhomogeneities can be seen as errors in $I_{\rm OT}$, as well as its first and second derivatives, evaluated at the $N$ target tweezers positions $(x_m,y_m)$. Their origin can be first attributed to the remaining discrepancies between the calculated $|A_{\rm SLM}|^2$ and $I_0$, which generate a kind of `speckle' interfering with the ideal tweezers array. In addition, we can also list many potential imperfections in the optical system that would also produce inhomogeneities: the finite pixel size of the SLM, as well pixel cross-talk and calibration errors of the displayed phase; uncorrected wavefront aberrations; diffraction caused by dust on optical surfaces or beam clipping on too small relay elements; interferences from multiple reflections on imperfect AR coatings... 

Fortunately, these inhomogeneities can be efficiently remove with a general optimization procedure.
First, a measurement estimates the individual deviations for each tweezers in a given observable (intensity, position or shape). 
Then, the pre-factors $f_m$ in Eq.(\ref{eq:interference_tweezers}), that introduce additional degrees of freedom per tweezers, are modified to correct the individual deviations.
This is repeated until the inhomogeneities are not detectable anymore within the measurement noise level. 
An important point of this procedure is to not modify anymore the choice of phases $\theta_m$, after the initial GS optimization. Indeed, as mentioned in the previous paragraph, these phases determine part of the inhomogeneities. Running again the GS optimization to keep optimizing these phases for high diffraction efficiency, while also trying to homogeneize the hologram, does not converge well, as discussed in Ref.~\cite{KimLukinOptica2019}. 


In the rest of this work, we put this homogenization procedure into practice. The experiments are performed on an existing setup previously described in Ref.~\cite{Chew2022Natphot}. The optical tweezers arrays are created by a $\lambda =850$~nm laser, reflected by an SLM (Hamamatsu-Photonics) and focused by a 0.75 NA objective lens (Special Optics). Ultracold $^{87}$Rb atoms are trapped in the tweezers array and used to perform high-precision \textit{in-situ} measurements of the tweezers. The optical aberrations introduced by the SLM itself and the rest of the optical system are already corrected with the hologram, and not discussed here. The tweezers intensities are homogenized in Section~\ref{sec:intensity}, the shapes in Section~\ref{sec:level4}, and the positions in Section~\ref{sec:position}.


\section{\label{sec:intensity}INTENSITY HOMOGENIZATION}
Homogenizing the intensity of tweezers across an array is important for a variety of applications. Without any optimization, the intensity of holographic tweezers can vary by 10~\% to 20~\% (rms). While such inhomogeneities would not affect much basic cooling and trapping, it become detrimental to more advanced applications.
For example, for
Rydberg dressing~\cite{Raman_Christian_2021} or for optical clocks in tweezers~\cite{Madjarov_endres_clock_2019,Matthew_Kaufman_clock_science_2019}, where a narrow transition is addressed, a level of 1~\% of inhomogeneity was shown to be limiting. Even more stringent, for exploring the tunneling dynamics of atoms between tweezers, much better than 0.5~\% was needed~\cite{kaufman_2014_science,Bakr_2022_prl}.
In this section, we investigate the intensity homogenization of optical tweezers, leveraging the concept of the elementary holograms we previously discussed. We perform a feedback optimization loop~\cite{GerchbergSaxton1972}, utilizing a Ramsey interferometer for precise \textit{in situ} measurement of the optical tweezers' intensity~\cite{ramsey_trap_shape_2021}. The performance of our optimization method is then demonstrated and discussed.

\begin{figure}[t]
    \centering
    \includegraphics[width= \columnwidth]{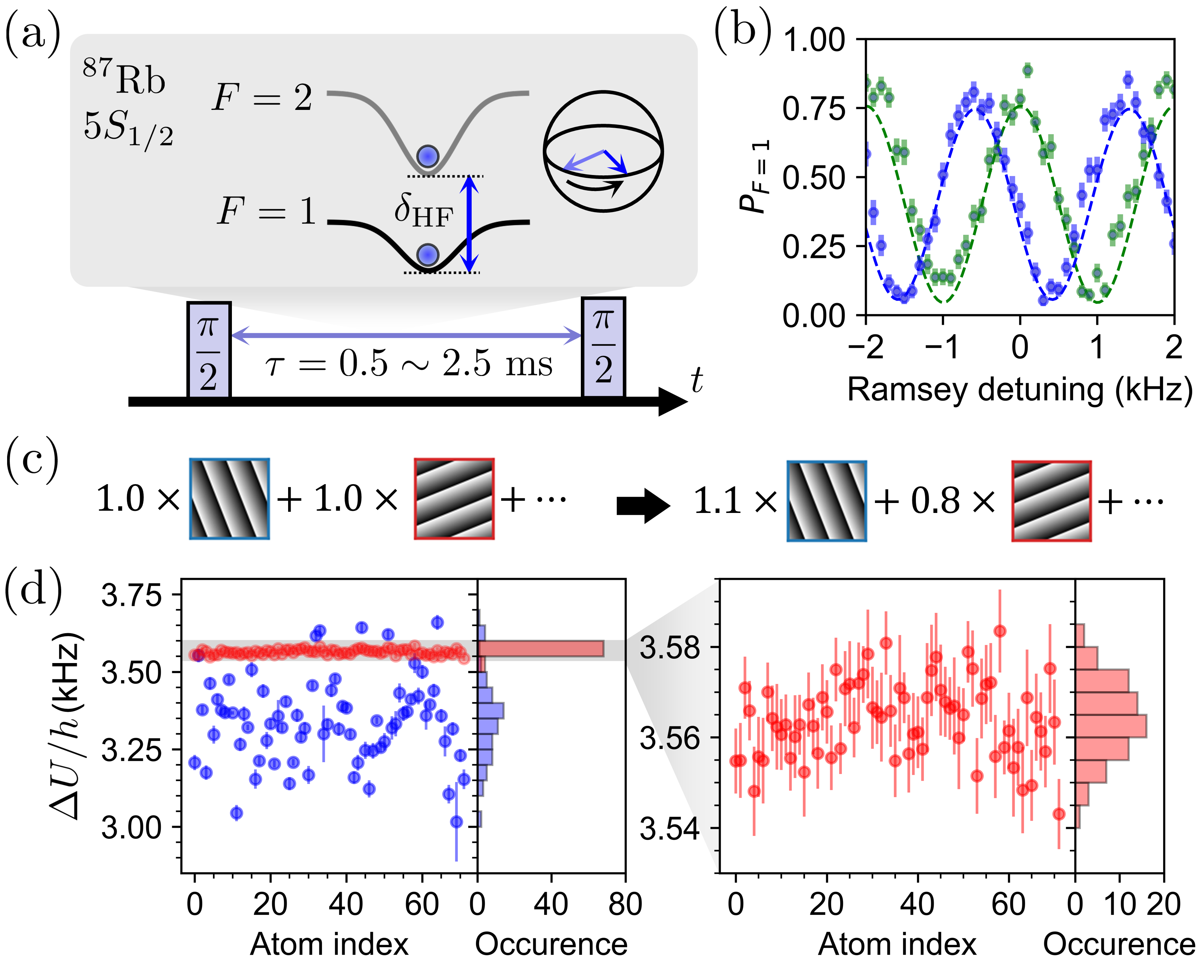}
    \caption{\textbf{Intensity homogenization with Ramsey interferometry}. (a) The tweezers intensity of a given site induces a differential light-shift on the clock transition of the trapped $^{87}$Rb atom. The latter is measured by Ramsey interferometry. (b) Ramsey fringes for two different atoms. The shift is caused by a different tweezers intensity. (c) Schematic showing the amplitude adjustment of each elementary hologram. (d) Differential light shift of the individual atom before (left) and after (right) homogenization, together with histograms. After the homogenization procedure, the intensity inhomogeneity decreased from 4~\% to 0.3~\%.}
    \label{MZI_discriminator}
\end{figure}

We quickly review how a hologram is optimized to homogenize the tweezers depth. One starts with Eq.~\eqref{eq:interference_tweezers} and recognizes that the amplitude of each elementary holograms $f_m(\mathbf{r}) = a_m^n$ can be adjusted in a feedback optimization ($m$ is the tweezers index, $n$ is the iteration of the optimization), as illustrated in Fig.~\ref{MZI_discriminator}(c). The tweezers intensity $I_m^n$ is estimated and the hologram parameters adjusted following: 
\begin{equation}
a_m^{n+1} = a_m^{n} \sqrt{\frac{\langle I \rangle}{I_{m}^n}}.
\label{eq:intensity_feedback}
\end{equation}
This procedure, which is often referred to as the \textit{weighted} GS algorithm, is already well adopted by the community~\cite{LeonardoTweezers,Nogrette2014}. An important subtlety, described in the previous section and repeated here, is to fix the choice of phases $\theta_m$ generated by the GS routine to allow an efficient convergence~\cite{KimLukinOptica2019}, instead of optimizing $a_m$ and $\theta_m$ together. 
The hologram would typically be obtained with a first stage of optimization based on numerical calculation of the diffracted intensities (which is fast), and then with a second stage using precise measurements (time-consuming). It is encouraged to already correct for known sources of inhomogeneities in the first stage, such as decreasing diffraction efficiency away from the optical axis (caused by the finite SLM pixel size), or decreasing the Strehl ratio of the optical system. This allows to produce tweezers with a homogeneity at the level of 5~\%, even for large-scale arrays~\cite{manetsch2024tweezer,pichard2024rearrangement_pasqal}, before moving to the second stage based on precise in-situ measurements.

Now we investigate how to precisely measure the intensity of each tweezers. 
The simplest option is to use a diagnostic camera re-imaging the tweezers~\cite{Nogrette2014}. However, it always deviates by typically a few percent from the intensity at the tweezers plane, due to imperfect optical systems (fringes caused by the camera cover glass, aberrations, ...).  
Therefore, \textit{in situ} measurement of the tweezers intensity utilizing atoms as probes are preferred. Previous attempts have demonstrated the use of atom fluorescence \cite{TamuraTweezers}, atom loading probability~\cite{tweezerbrowaeys2021}, atom spectroscopy \cite{Cooper_Endres_PRX_2018}, and atom density profiles \cite{Bakr_2022_prl} to reconstruct the tweezers depth. However, these approaches face certain limitations. They either achieve only a few percent of precision due to the interrogation of broad spectral lines~\cite{TamuraTweezers,tweezerbrowaeys2021,Cooper_Endres_PRX_2018}, or are too application-specific~\cite{Bakr_2022_prl}. 
In this work, we rather interrogate the $5S_{1/2}$ hyperfine clock transition of our $^{87}$Rb trapped atoms, which offers a much better ratio of coherence time ($T_2^\star \sim 10$~ms) to sensitivity. The latter is given by the differential light shift: 
\begin{equation}
\Delta U = \eta U \approx \frac{\delta_{\rm HF}}{\delta} U
\end{equation}
with $\delta_{\rm HF} = 6.8\,{\rm GHz}$ the hyperfine splitting, $\delta = 30\,{\rm THz}$ the tweezers detuning from the $5P$ state, and $\eta = 2.3\times10^{-4}$ the ratio between the light-shift $U$ of the $5S$ state and the differential light-shift $\Delta U$ between its two hyperfine states.

The measurement of $\Delta U$, and its variation from tweezers to tweezers, is then performed by Ramsey interferometry~\cite{ramsey_trap_shape_2021}, as shown in Fig.~\ref{MZI_discriminator}(a). The resulting fringes for two atoms are shown in Fig.~\ref{MZI_discriminator}(b) for an interrogation time of 0.5~ms, clearly showing a variation of $\Delta U/ h$ by 0.7~kHz, to be compared to a mean value of $\sim 3.5$~kHz. This information is then fed to the optimization routine through Eq.~(\ref{eq:intensity_feedback}).
The optimization loop is repeated with increasing interrogation times, from 0.5 ms up to 2.5 ms, as the inhomogeneity decreases, and higher precision is required. Starting from an array with 4~\% inhomogeneity (140~Hz), we could suppress it down to 0.3~\% (11~Hz) across an array of 72 atoms in 3 minutes of experimental time, as illustrated in Fig.~\ref{MZI_discriminator}(d). 

We conclude with prospects for further improvement. 
In this work, our inhomogeneity level was limited by the measurement precision of 10~Hz, as shown by the error bars in Fig.~\ref{MZI_discriminator}(d). This limitation is currently set by the finite $T_2^{*}$ time, causing the Ramsey contrast to decay, which we attribute to phase flicker noise from the SLM~\cite{Yang_phase_flicker_2020} and intensity noise of the trap laser. Future works along that direction could help in extending the $T_2^{*}$ time.
Additionally, we also verified that the SLM finite phase resolution (8 bits) was not yet a limiting factor. For this, we intentionally  reduced the SLM resolution from 8 to 7 and 6 bits per pixel, but observed no impact on the final inhomogeneity reached by the optimization procedure (as also supported by numerical simulations).
We thus expect that arrays with inhomogeneity even lower than the 0.3~\% achieved here could be created if a very demanding application requires it. 

\section{\label{sec:level4}Shape homogenization}

The shape of the optical tweezers is another important parameter, as it influences the motion of atoms in the traps. This shape is parameterized by a set of three trapping frequencies $\omega_{x,y,z}$, related to the curvature of the potential along the three normal directions. In the standard approximation of a perfect Gaussian beam, we obtain:
\begin{equation}
\omega_{x,y} = \sqrt{\frac{4U}{mw^2}}; \, \omega_z =  \sqrt{\frac{2U}{mz_r^2}}
\label{eq:trap_freq_waist}
\end{equation}
where $w$ ($z_r$) represents the waist (Rayleigh length), $U$ the trap depth, and $m$ the mass of the atom. 
When considering an array of tweezers, it is not uncommon to observe an inhomogeneity of 10~\% (even after having homogenized the trap depth). For standard manipulation of the atom, such as atom re-arrangement~\cite{Barredo2016Nov,Endres2016Nov,Kim2016Oct}, molecule assembly~\cite{tweezer_molecule_kkNI_2018,merge_Cornish_2023}, and cooling to the motional ground-state~\cite{KKN_PRA_ramansideband_2018, Lu2024Mar_raman,Raman_Christian_2021,Chew2022Natphot}, this inhomogeneity is typically overcome efficiently by the use of robust, adiabatic, manipulation. However, more advanced use-cases require a finer control. For example, to explore spin-motion coupling between ground-state~\cite{spin_motion_zhan_2024}  or between Rydberg atoms~\cite{bharti2023strong, Chew2022Natphot,spin_phonon_igor_2024}, a control to better than 10~\% is beneficial. For the most stringent applications, where quantum states of motion are manipulated, such as rotational Laughlin states~\cite{lunt2024realization}, squeezed states~\cite{Inprep}, cat states~\cite{spin_motion_lkb_2023}, spin-motion entangled state~\cite{scholl2023erasurecooling}, or tunneling between tweezers~\cite{kaufman_2014_science,Bakr_2022_prl}, a precision of better than 1~\% in inhomogeneity and anisotropy is required. 

\begin{figure}[t]
    \centering
    \includegraphics[width= \columnwidth]{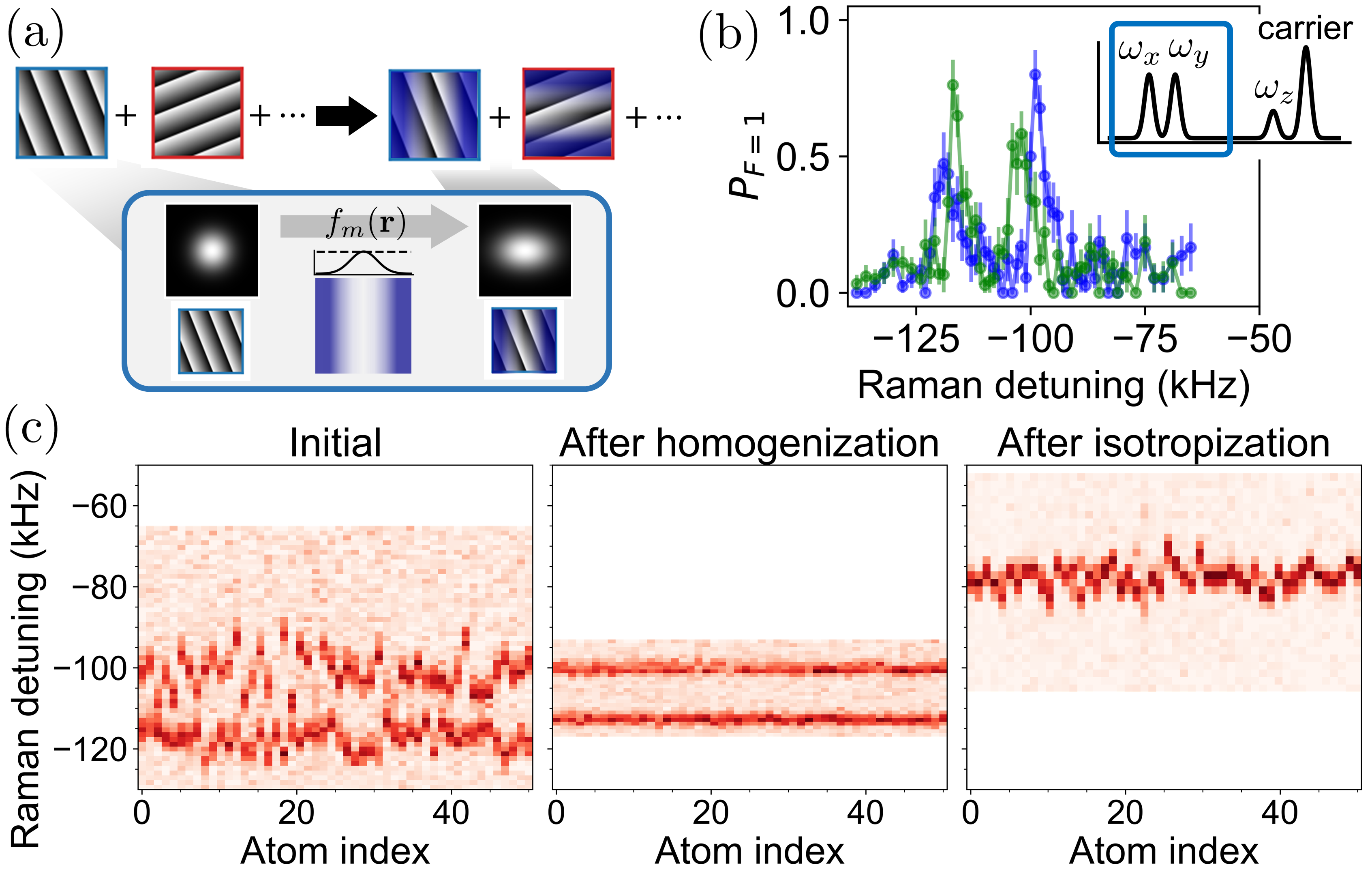}
    \caption{\textbf{Shape homogenization by apodization}. (a) Schematic showing the optical apodization with elliptical Gaussian mask applied on the individual elementary holograms. The apodizing parameters are optimized tweezers-by-tweezers to correct their individual trapping frequencies. (b) Raman sideband spectroscopy spectrum of two different atoms. The two peaks are located at the trapping frequencies on the $x$ and $y$ directions. The measured frequencies $\omega_x$ and $\omega_y$ are compared to the array average and fedback to the hologram for correction. (c) Raman sideband spectroscopy of the 50-atoms array before homogenization (left), after homogenization (middle), and after isotropization (right).}
    \label{fig:shape}
\end{figure}

Having recognized the importance of controlling the tweezers shape for quantum engineering with trapped atoms, we now move to the characterization of the trapping frequencies, with a focus on \textit{anisotropy} in a single tweezers and \textit{inhomogeneity} over an array. 
Measurement can be done by parametric heating or excitation of breathing mode~\cite{Sortais2007}, but the best frequency resolution is obtained by Raman sideband spectroscopy~\cite{Kaufman_raman_2012,Thompson_raman_2013}. Figure~\ref{fig:shape}(b) shows a sketch of the expected spectrum with 3 peaks centered at each trapping frequency. Because of the Gaussian beam profile, the tweezers is strongly anisotropic with a much weaker curvature along the optical axis of $\omega_{z} = 2\pi \times 25$~kHz than in the radial (focal) plane $\omega_{x,y}> 2\pi \times 100$~kHz. 
Interestingly, our tweezers also display an anisotropy of 20~\% between the $x$ and $y$ direction caused by the choice of a linearly polarized tweezers (along $x$) breaking the radial symmetry due to vector diffraction effect~\cite{Richards_Wolf_highna}, which are stronger here than in typical tweezers setup due to the choice of a very high-NA objective (NA = 0.75). 
Finally, we look at the inhomogeneity from tweezers to tweezers. In Fig.~\ref{fig:shape}(b), spectra from two different atoms are compared, displaying different radial trapping frequencies. This is even clearer in Fig.~\ref{fig:shape}(c) showing an inhomogeneity of around 10~\% over an array of 50 tweezers. 

We choose to focus solely on the shape on the radial plane, as the motion along the optical axis affects slower dynamics and can thus be ignored in first order. If the correction of the anisotropy between radial and axial axes were to be needed, an optical lattice can be applied along the optical axis to increase the confinement along that direction~\cite{Young2020}. 
To modify the shape of tweezers, we use the concept of optical apodization~\cite{goodman2005introduction}. As illustrated in Fig.~\ref{fig:shape}(a), by \textit{obscuring} part of the hologram in a specific direction, we can extend the optical tweezers along that direction. Such apodization can be applied and tuned for each individual tweezers allowing to obtain a radially isotropic and homogeneous tweezers array. 
To incorporate this idea in the Eq.~(\ref{eq:interference_tweezers}) defining the hologram, we now allow for a slowly-varying amplitude $f_m$ as: 
\begin{equation}
f_m(\mathbf{r}) =a_m \exp{\left[ -\frac{\tilde{x}^2}{X_m^2} -\frac{\tilde{y}^2}{Y_m^2}\right]}
\end{equation}
which introduces two additional degrees of freedom per tweezers for optimization, $X_m$ and $Y_m$, the apodization lengths along the $\tilde{x}$ and $\tilde{y}$ direction, which are now added to the amplitude-tuning parameter $a_m$. One more tuning knob has also been implicitly added through the possibility to rotate the elliptical Gaussian mask with natural axis ($\tilde{x},\tilde{y}$) around the SLM axis ($x, y$), but was kept fixed here~\cite{angle_shape}.




Having now a set of four parameters per tweezers ($\theta$, $a$, $X$, $Y$), we present our strategy to optimize them together. In Section~\ref{sec:level2}, we already concluded that $\theta$ is optimized first by the GS method and then kept fixed, so we can focus on the remaining three to adjust the trap depth $U$ and radial frequencies $\omega_{x,y}$. As seen in Eq.~(\ref{eq:trap_freq_waist}), there is an obvious coupling between the three trap parameters to be considered. In addition, while we previously assumed a linear relation between the tweezers amplitude $\sqrt{U}$ and the parameter $a$ in Eq.~(\ref{eq:intensity_feedback}), this is more complicated for the relation between apodization lengths $X,Y$ and trap frequencies $\omega_{x,y}$. Indeed, the latter also depends on the choice of beam waist illuminating the SLM, the objective diameter and aberrations. We encapsulated the response of the tweezers ($U, \omega_x, \omega_y$) to each parameter ($a, X, Y$) into a $3 \times 3$ matrix $C$ which was calibrated experimentally. This ensures an efficient convergence of the optimization routine. The optimization of the hologram then proceeds as follows: we have a first round of trap depth optimization (described in Section~\ref{sec:intensity}), then measure the individual trapping frequencies and generate an error signal $\Delta \omega_{x,y}$ for each tweezers which are fed into:
\begin{equation}
\begin{pmatrix} a \\ X \\ Y \end{pmatrix}_{n+1} = \begin{pmatrix} a \\ X \\ Y \end{pmatrix}_n + 
C^{-1} \begin{pmatrix} 0 \\ \Delta\omega_x \\ \Delta\omega_y \end{pmatrix}_n
\end{equation}
to generate the new round of tweezers parameters $(a,X,Y)_{n+1}$ where $n$ is the index of the shape optimization loop. We then go back to the depth optimization and repeat typically 5 times, which takes around 40 minutes of experimental time. 



We now present the performance of the optimization routine. 
We start with an array of tweezers with an inhomogeneity of 5~\% for both $x$ and $y$ trapping frequencies, as seen in Fig.~\ref{fig:shape}(c,~left). Following the optimization process described above successfully brings the inhomogeneity to below 0.5~\%, as clearly apparent in Fig.~\ref{fig:shape}(c,~middle). To our knowledge, this is the first time to achieve such a high shape homogeneity in a holographic tweezers array. As a second demonstration of shape optimization, we also corrected the anisotropy of the trap in the two radial $(x,y)$ directions. Here, we applied a global correction, \textit{i.e.}, the apodization mask is the same for all traps, and optimized its ellipticity and orientation to minimize the anisotropy. The result is shown in Fig.~\ref{fig:shape}(c,~right), where a double peak is now indistinguishable within the finite width of the sideband resonance (FWHM: 5 kHz). This puts a higher-bound on the residual anisotropy of less than 5~\%, an improvement of more than 4 compared to the initial hologram.

We have successfully demonstrated how to optimize the shape of holographic tweezers over an extended array. The technique could be pushed further, by (1) improving the measurement resolution and (2) performing simultaneously the homogenization and radial isotropization. For point (1), the resolution is currently set by the probe pulse duration (and sweep range, since we use hyperbolic-secant adiabatic pulses), which could be improved immediately. We would then be limited by the coherence time of motion and hyperfine spin; the later being shorter with a $T_2^\star = 10$~ms coherence time (on the hyperfine clock transition, limited by fast noise of the trap depth). One could implement a more advanced scheme, where a Ramsey sequence is interleaved with spin-echoes to reach the ultimate limit set by the motional state coherence time (hundreds of milliseconds, from scattering of the tweezers photons). For point (2), this would require to identify the directions of the motional modes, which was easily done here thanks to the large anisotropy, but is more challenging for quasi-isotropic tweezers. A simple upgrade of the measurement technique would allow such identification by performing the Raman sideband spectroscopy along two independent directions, and then reconstructing the physical orientation of the motional resonances from the two spectra. 



\section{POSITION HOMOGENIZATION}\label{sec:position}

The position of optical tweezers, defined as the point of highest intensity, is crucial as it determines the center of the confined atomic wavefunction. Accurate positioning is particularly important when considering interactions with other atoms~\cite{Chew2022Natphot,Barredo2015_prl,kaufman_2014_science}, macroscopic objects~\cite{thompson_lukin_science_2013, Deist_2022, menon2023integrated, Ferri_2022}, or the electromagnetic field \cite{Tamura_2020,Yan_Dan_super_2023,Shaw_endres_NPHY_2024}. A reasonable target for the precision of tweezers positioning would be to reach below the quantum uncertainty in the motional ground-state of the tweezers: $\sqrt{\hbar/2m\omega} \sim 20 - 50$~nm. 
In this section, we focus on measuring and correcting the tweezers position along the optical axis, where we have the least information from standard fluorescence imaging. The tweezers fine positioning is performed with a Fresnel lens for each optical tweezers. We measure the atom position with an \textit{optical ruler}, and finally perform the position homogenization with a feedback on the hologram. The performance of our measurement and feedback results are shown and discussed.

\begin{figure}
    \centering
    \includegraphics[width=1.0\linewidth]{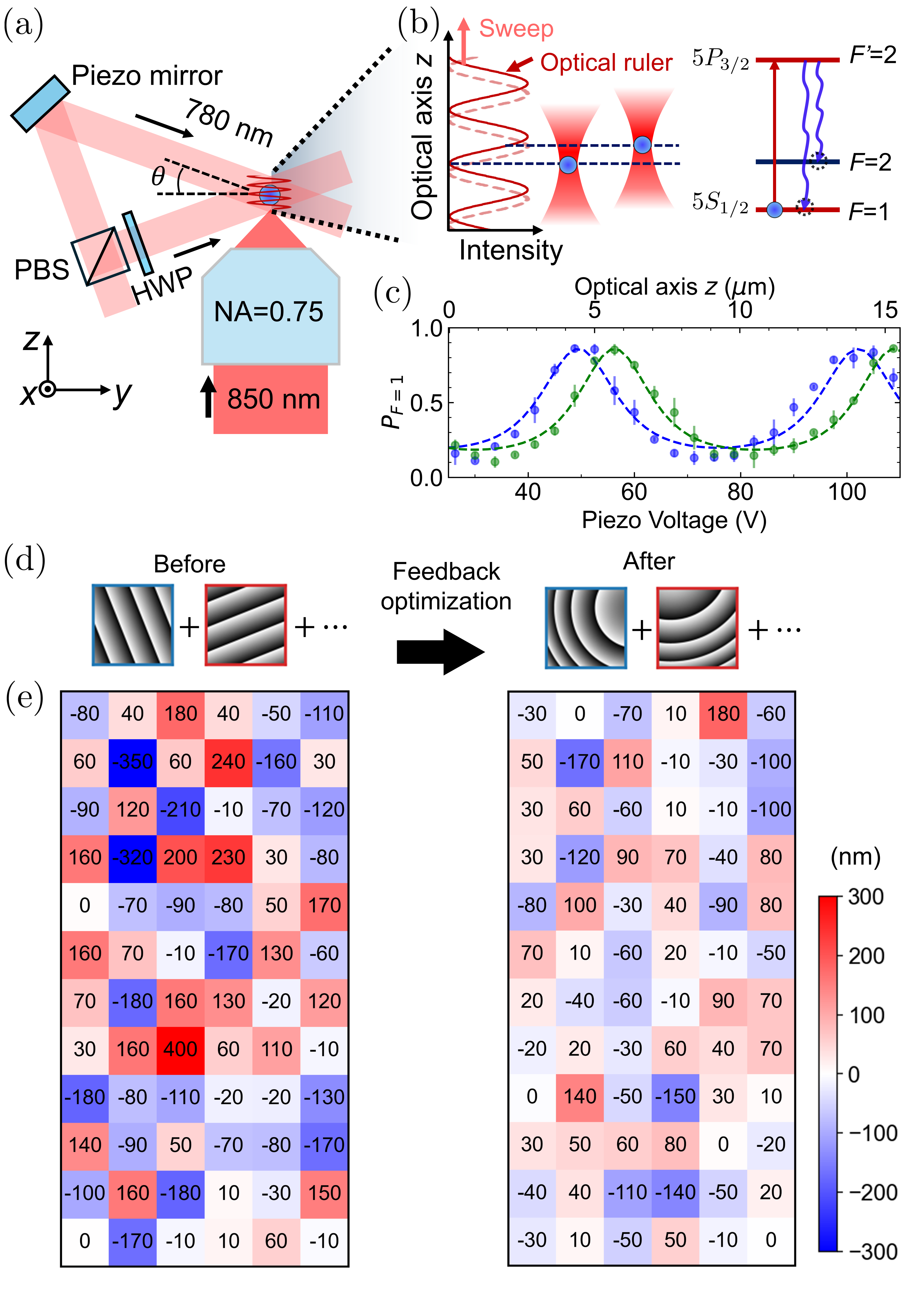}
    \caption{\textbf{Position homogenization with an optical ruler}. (a) The optical ruler is created by interfering two laser beams on the atomic array. (b) The $z$-position of the atom is measured by scanning the optical ruler which is resonant with the $\ket{\mathrm{5}S_{1/2}, F=1}  \rightarrow \ket{\mathrm{5}P_{3/2}, F'=2}$ transition. (c) Experimental data for two atoms with their position with respect to the ruler extracted from a numerical fit. (d) Fresnel lenses are employed to correct the SLM hologram. (e) The position of each atom is indicated relative to the average atomic plane. The standard deviation over the array is reduced from 130~nm (left) to 70~nm (right) after applying two loops of feedback optimization of the hologram.}
    \label{figoptical system}
\end{figure}

We set the position of each optical tweezers by adjusting the elementary hologram as described in equation (\ref{eq:interference_tweezers}). These adjustments involve gratings characterized by the wavevector $\mathbf{k_m}= \frac{2\pi M}{\lambda F}(x_m,y_m)$ and a new Fresnel lens phase profile $\frac{\pi M^2}{\lambda F^2}z_m(x^2 + y^2)$ \cite{Barredo2018}, adding an individual tunable defocus to each tweezers~[Fig.~\ref{figoptical system}(d)]. By tuning the parameters $(x_m, y_m, z_m)$, where $x_m$ and $y_m$ are the coordinates in the focal plane and $z_m$ is the position along the optical axis, we can achieve precise control of the optical tweezers. The positioning resolution is set by the SLM finite number of pixels as well as by the phase resolution at each pixel (here, $2\pi/256$), which we estimate to be 5~nm in the radial direction and 15~nm along the optical axis, well below our target. Here, we will only tune the tweezers position along the $z$ optical axis. 

We now consider how to accurately measure the position of optical tweezers. Such measurements are typically conducted by imaging the tweezers on a diagnostic camera or by using fluorescence image of single atoms trapped in the optical tweezers. With these approaches, the precision along the optical axis would be limited to a fraction of the Rayleigh length at the level of a few hundreds nanometers. Additionally, such ex-situ measurement is influenced by imperfections in the optical system (for example, field curvature) and uncertainties in the magnification ratio $M$. To overcome these limitations, we prefer to measure the atom position with an \textit{in-situ optical ruler}, also introduced in previous research~\cite{McDonald_2019}. This method measures the position of atoms with respect to an interference pattern applied on the atomic array. We will see that it offers both improved resolution (as the ruler period can be set to as low as half the wavelength), and can also be calibrated precisely.

The optical ruler is created inside the vacuum chamber through the interference of two beams, generating a lattice structure depicted in Fig.~\ref{figoptical system}(a). The lattice constant $a = {\rm 9.6(1)~\mu m}$ is confirmed with a camera, determined by the relative angle $2\theta \simeq 4.6^\circ$ between the two lattice beams, limited by the numerical aperture of the vacuum chamber along the chosen direction. The optical ruler is tuned to resonance with the $\ket{\mathrm{5}S_{1/2}, F=1}$ to $\ket{\mathrm{5}P_{3/2}, F'=2}$ transition, as shown in Fig.~\ref{figoptical system}(b). Atoms initially prepared in $\ket{\mathrm{5}S_{1/2}, F=1}$  are pumped to $\ket{\mathrm{5}S_{1/2}, F=2}$ by the optical ruler, with the probability $P_{F=1}$ to remain in the initial state now depending on the ruler intensity at the atoms' position. As the phase of the optical ruler is scanned with a piezo mirror on one of the beam paths, the atom position in the ruler is revealed. Figure~\ref{figoptical system}(c) shows a resulting dataset for two atoms at different heights, which can then be extracted by a numerical fit with a precision of 50 nm. This precision of 0.5~\% of the lattice period $a$ (allowed by saturating the transition~\cite{McDonald_2019}) reaches the target defined earlier. Further precision would require a smaller spacing, which could be obtained with a retro-reflecting beam along the $z$-axis. 

We now demonstrate measurement and homogenization over a large array of 72 atoms with an inter-atomic separation of 5~${\rm \mu m}$. After measuring the atom position with respect to the optical ruler, we then extract the average atomic plane and the deviation of each atom from it, which is then shown at the 2D map of Fig.~\ref{figoptical system}(e, left).
Initially, the standard deviation along the optical axis is 130 nm, which is clearly larger than the measurement precision. We then improve the hologram, with two rounds of feedback optimization where the Fresnel lens of each tweezers is fine tuned (see Fig.~\ref{figoptical system}(d)), and can finally reduce the standard deviation of the positions to 70~nm, which is shown in Fig.~\ref{figoptical system}(e,~right). Currently, our position optimization is limited by the 50~nm measurement uncertainty.

In conclusion, we demonstrated a precise position measurement of the tweezers with an optical ruler, reaching a 50~nm measurement error, which is more accurate than using fluorescence images. Our method effectively enhances optical tweezers position control by roughly a factor of 2, reducing the standard deviation of atom positions from 130~nm to 70~nm, almost at the level of the quantum fluctuation of the atom in the tweezers. Further refinements, such as employing a smaller lattice constant for the optical ruler to reduce the measurement error, and repeating the procedure for each direction, could lead to 3D positioning with nanometer precision. 

\section{CONCLUSION AND OUTLOOK}\label{sec:summary}

In this study, we systematically tackled the challenges associated with optimizing optical tweezers arrays generated by an SLM, focusing on the homogenization of intensity, shape, and position. Visualizing the hologram as the sum of elementary ones which are then individually tailored (apodization, Fresnel lens, ...) allows to easily parametrize the hologram optimization. By then implementing refined measurement techniques such as Ramsey interferometry, Raman sideband spectroscopy, and optical ruler, we could input precise information on the array in the feedback optimization of the hologram. As a result, the inhomogeneity over an atomic array of 50-100 atoms was reduced in intensity to 0.3~\% (rms), in shape to 0.5~\%, and in position to 70~nm along the optical axis. 

We believe that the techniques presented here are scalable with atom numbers, as the measurement is performed in parallel, and the SLM offers enough degrees of freedom with millions of pixels each with high phase resolution. The precision achieved here could be further improved, by lowering the measurement uncertainties, if future applications require it.  
Fine tuning each tweezers can be even pushed further by introducing higher Zernike polynomials for tweezers-by-tweezers aberration control (here we corrected only the defocus term in Section~\ref{sec:position}). It is also possible to remodel individual tweezers into bottle beam traps~\cite{Barredo_bottle_2020} or high-order Laguerre–Gauss modes~\cite{kimble_2020_tweezer} with fancier individual phase masks. Tweezers-by-tweezers control mentioned above could be combined with global aberration correction of the optical system~\cite{hill2024opticalphaseaberrationcorrection_Johim_2024}, which may further improve our tweezers homogeneity.

We envision that the precise controls of the atomic array would be crucial when the interaction between Rydberg atoms is directly used for quantum simulation or computation with Rydberg atoms~\cite{Browaeys2020Feb_review, Chew2022Natphot, Chen2023, Ahn2024} (\textit{i.e.}, not in the Rydberg blockade regime), for Hubbard physics in tweezers~\cite{Kaden_hubbard_2024}, super-radiance in ordered arrays~\cite{Robicheaux_collective_2016,Masson_radiance_2022}, tweezers interferometers~\cite{tweezer_inter_2023}, or interfacing tweezers with optical lattice~\cite{Aaron_kaufman_2022,tao2023highfidelity}, or nanophotonic waveguides~\cite{Tamura_chen_2023,menon2023integrated,Luan_kimble_2020,Tamara_lukin_2021}. 

\begin{acknowledgments}
M.P. thanks R. Martin for helpful discussions about shape homogenization. This work was supported by MEXT Quantum Leap Flagship Program (MEXT Q-LEAP) JPMXS0118069021, JSPS Grant-in-Aid for Specially Promoted Research Grant No. 16H06289 and JST Moonshot R\&D Program Grant Number JPMJMS2269.
\end{acknowledgments}
Y.T.C. and M.P. contributed equally to this work.


\begin{thebibliography}{84}
\bibitem [{ang()}]{angle_shape}%
  \BibitemOpen
  \href@noop {} {}\bibinfo {note} {We globally optimized this rotation angle for all tweezers (to account for a slight rotation caused by a periscope), but it could also be fine-tuned tweezers-by-tweezers.}\BibitemShut {Stop}%
\end{thebibliography}
\end{document}